\documentclass[conference]{IEEEtran}
\IEEEoverridecommandlockouts
\usepackage{amsmath,amssymb,amsfonts}
\usepackage{algorithmic}
\usepackage{graphicx}
\usepackage{textcomp}
\usepackage{xcolor}
\usepackage{amsmath,graphicx}
\usepackage{multirow}
\usepackage{makecell}
\usepackage{caption}
\usepackage{lineno,hyperref}
\usepackage{subcaption}
\usepackage{wrapfig}
\usepackage{url}
\usepackage{hyperref}
\usepackage{cite}
\def\BibTeX{{\rm B\kern-.05em{\sc i\kern-.025em b}\kern-.08em
    T\kern-.1667em\lower.7ex\hbox{E}\kern-.125emX}}
\begin{document}

\title{Mutual Evidential Deep Learning for Medical Image Segmentation\\
\thanks{This work is supported by National Key R$\&$D Program of China (2023YFC3502900), National Natural Science Foundation of China (granted No. 62192731) and National Key R$\&$D Program of China (2021YFF1201100). Corresponding author: Zhi Jin.}
}

\author{\IEEEauthorblockN{1\textsuperscript{st} Yuanpeng He}
\IEEEauthorblockA{\textit{Key Laboratory of High Confidence} \\
	\textit{Software Technologies (MOE)} \\
	\textit{School of Computer Science} \\
	\textit{Peking University},
	Beijng, China \\
	heyuanpeng@stu.pku.edu.cn}
\and
\IEEEauthorblockN{2\textsuperscript{nd} Yali Bi}
\IEEEauthorblockA{\textit{College of Computer and Information Science} \\
	\textit{School of Software} \\
\textit{Southwest University},
Chongqing, China \\
biyali812@outlook.com}
\and
\IEEEauthorblockN{3\textsuperscript{rd} Lijian Li}
\IEEEauthorblockA{\textit{Department of Computer and} \\
	\textit{Information Science} \\
\textit{University of Macau}, Macau, China  \\
mc35305@umac.mo}
\and
\IEEEauthorblockN{4\textsuperscript{th} Chi-Man Pun}
\IEEEauthorblockA{\textit{Department of Computer and} \\
	\textit{Information Science} \\
\textit{University of Macau}, Macau, China  \\
cmpun@um.edu.mo}
\and
\IEEEauthorblockN{5\textsuperscript{th} Wenpin Jiao}
\IEEEauthorblockA{\textit{Key Laboratory of High Confidence} \\
	\textit{Software Technologies (MOE)} \\
	\textit{School of Computer Science} \\
	\textit{Peking University}, Beijng, China \\
	jwp@pku.edu.cn}
\and
\IEEEauthorblockN{6\textsuperscript{th} Zhi Jin$^{*}$}
\IEEEauthorblockA{\textit{Key Laboratory of High Confidence} \\
	\textit{Software Technologies (MOE)} \\
	\textit{School of Computer Science} \\
	\textit{Peking University},	Beijng, China \\
	zhijin@pku.edu.cn}
}

\maketitle

\begin{abstract}
Existing semi-supervised medical segmentation co-learning frameworks have realized that model performance can be diminished by the biases in model recognition caused by low-quality pseudo-labels. Due to the averaging nature of their pseudo-label integration strategy, they fail to explore the reliability of pseudo-labels from different sources. In this paper, we propose a mutual evidential deep learning (MEDL) framework that offers a potentially viable solution for pseudo-label generation in semi-supervised learning from two perspectives. First, we introduce networks with different architectures to generate complementary evidence for unlabeled samples and adopt an improved class-aware evidential fusion to guide the confident synthesis of evidential predictions sourced from diverse architectural networks. Second, utilizing the uncertainty in the fused evidence, we design an asymptotic Fisher information-based evidential learning strategy. This strategy enables the model to initially focus on unlabeled samples with more reliable pseudo-labels, gradually shifting attention to samples with lower-quality pseudo-labels while avoiding over-penalization of mislabeled classes in high data uncertainty samples. Additionally, for labeled data, we continue to adopt an uncertainty-driven asymptotic learning strategy, gradually guiding the model to focus on challenging voxels. Extensive experiments on five mainstream datasets have demonstrated that MEDL achieves state-of-the-art performance.
\end{abstract}


\section{Introduction}
Medical image segmentation (MIS) \cite{DBLP:conf/cvpr/YuanXDCYQYYSCL023} is a formidable undertaking at the intersection of computer vision and healthcare, characterized by intricate challenges that demand meticulous consideration. These challenges encompass the inherent multimodality of medical data, evident disparities from natural images, and the difficulty of manual annotation \cite{najjar2023redefining}. Multimodality arises due to the diverse imaging modalities in use, such as CT scans and MRI, each marked by unique and complex image attributes. The substantial divergence from natural images is evident in the wide range of pixel values, the prevalence of artifacts and noise, and the lack of well-defined object boundaries, all of which significantly complicate conventional segmentation methods. Furthermore, high-quality annotated medical image datasets are notably constrained in scale, rendering semi-supervised learning a pragmatic and scalable solution for addressing the intrinsic challenges to MIS.

The current landscape of MIS methods can be broadly categorized into two main approaches. The first emphasizes uncertainty estimation to improve model robustness and prediction quality \cite{tang2022unified}. In this context, uncertainty-aware methods \cite{catak2022uncertainty} play a pivotal role by quantifying the model’s confidence in its predictions, thus enabling reliable refinement and error correction. Many recent works further advance this line of research by integrating evidential theory into segmentation models \cite{he2024uncertainty, he2025co, he2024epl}, offering a probabilistic framework for modeling aleatoric and epistemic uncertainties. The second category leverages semi-supervised strategies by jointly training on labeled and unlabeled data \cite{wang2022semi}, often incorporating data augmentation, consistency regularization, and pseudo-labeling. Techniques such as prototype-based embedding refinement \cite{bi2025multi}, prototype consistency \cite{li2024efficient}, and federated knowledge transfer \cite{huang2025unitrans} have shown significant improvements in performance and generalization. Nevertheless, a persistent challenge across existing approaches is the effective mitigation of recognition biases and the optimal exploitation of uncertainty information. To address these issues, the mutual evidential deep learning (MEDL) framework \cite{he2024mutual} introduces an advanced methodology that combines evidential deep learning (EDL) with evidential fusion mechanisms \cite{he2024uncertainty, he2025co}. Specifically, MEDL is designed to model voxel-wise uncertainty and adjust the learning strategy per voxel by fusing evidence from two complementary networks, each with its own architectural bias. Such fusion draws inspiration from generalized evidence theory \cite{he2022mmget}, ordinal belief entropy \cite{he2023ordinal}, and conflict management methods in belief assignment \cite{he2021conflicting, he2022new, he2022ordinal}, enhancing both robustness and adaptability. In the MEDL framework, two segmentation networks are employed to generate evidential predictions on both labeled and unlabeled data. For the unlabeled portion, class-aware evidential fusion (CAEF) integrates predictions from both networks to produce reliable pseudo-labels and associated uncertainty estimates \cite{he2024epl, he2023tdqmf}. Reliability-based masking, inspired by entropy theory and evidential entropy \cite{he2022ordinal}, is applied to suppress noisy predictions. A progressive Fisher Information-based Evidential Learning (FIE) module then facilitates a curriculum-style training regime, emphasizing confident voxels and gradually incorporating uncertain ones. This design also benefits from recent studies on temporal evidence fusion in time series forecasting \cite{zhan2024time}, matrix-based decision methods \cite{he2021matrix}, and adaptive entropy-driven clustering \cite{li2025adaptive}.

Additionally, the MEDL pipeline borrows from attention-based weakly-supervised localization \cite{he2024generalized}, neural structure modeling \cite{li2022nndf}, and fuzzy forecasting principles \cite{zhan2023differential}, integrating them into a coherent evidential framework. The use of feature-reutilization networks \cite{he2024residual} further boosts segmentation accuracy by enhancing spatial contextual representation. Interestingly, MEDL’s iterative pseudo-label refinement and progressive training paradigm resemble self-debugging techniques in code generation systems \cite{chen2025revisit}, where a model generates intermediate outputs and continuously self-corrects through structured evaluations. Moreover, its ability to leverage heterogeneous predictions across modalities can be analogized to multi-temporal image classification frameworks \cite{xu2023spatio}, which address spatio-temporal variability across complex datasets.

In summary, the MEDL approach not only leverages the strengths of evidential modeling and uncertainty estimation but also inherits and expands upon recent developments in fuzzy systems, deep evidential learning, and semi-supervised segmentation. As such, it presents a promising direction for future research on robust, scalable, and uncertainty-aware medical image segmentation. The main contributions of this work are given as follows:
\begin{itemize}
	\item The proposed method designs CAEF to combine complementary evidences from two heterogeneous segmentation models. This technique enhances the confidence of labeled data predictions and generates reliable pseudo-labels by filtering out uncertain information.
	\item The method proposes an innovative strategy for asymptotic learning using uncertainty-based estimations. By integrating the evidential predictions from two networks and their built-in uncertainties, which guides the easy-to-hard and certain-to-uncertain learning process.
	\item Experiments strongly demonstrate the effectiveness of MEDL which far outperforms previous state of the arts.
\end{itemize}

\begin{figure*}
	\centering
	\setlength{\belowcaptionskip}{-15pt}
	\includegraphics[scale=0.223]{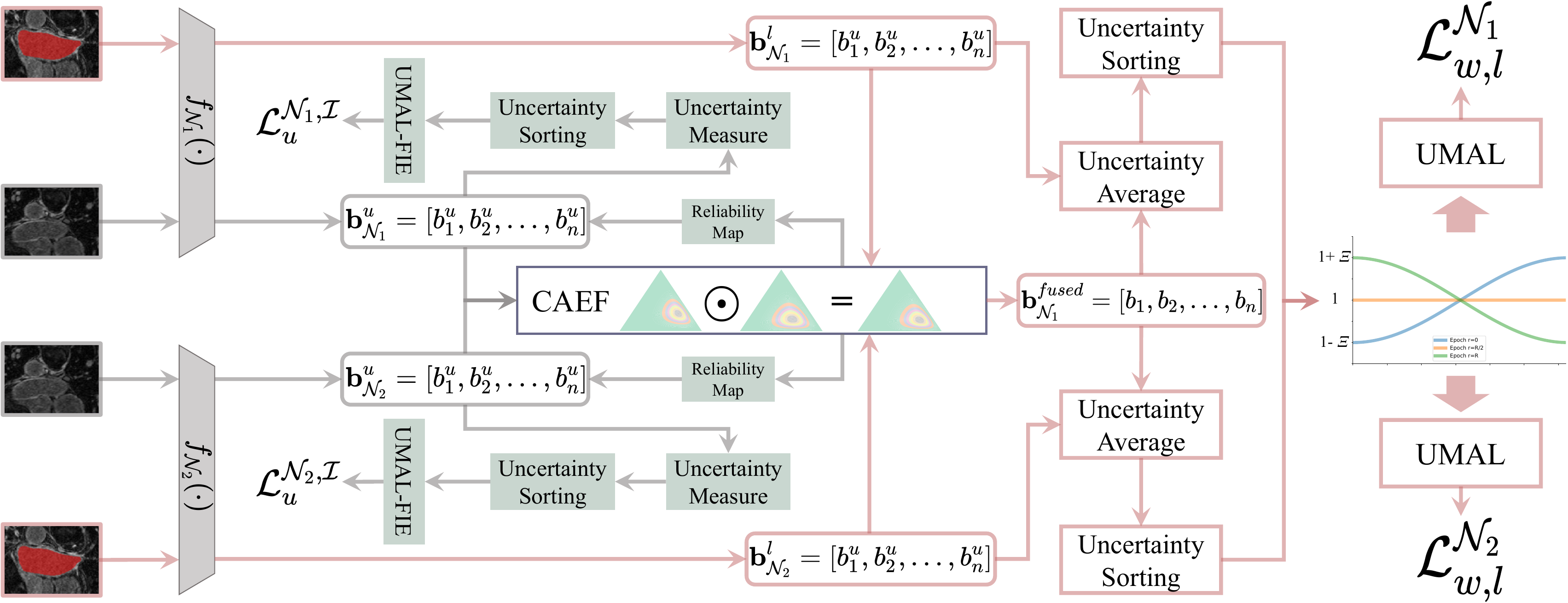}
	\caption{The overview of Mutual Evidential Deep Learning framework. We use two different segmentation models to predict evidence for labeled and unlabeled data. For unlabeled data, class-aware evidential fusion (CAEF) combines complementary evidence from the two models to generate pseudo-labels and reliability measures. Reliability measures mask the original evidential predictions from each network, and evidential deep learning (EDL) uncertainty with weighted average estimates rank each voxel, applying asymptotic Fisher information-based EDL (FIE). For labeled data, a similar strategy guides learning without the need for reliability-based masking and FIE ($\odot$ represents the fusion operation).}
	\label{fig:model}
\end{figure*}
\section{Method}
\label{sec:method}
The details of the proposed method are provided in the Figure \ref{fig:model}. Specifically, the mutual evidential deep learning framework consists of two networks, $\mathcal{N}_1(\Theta_1)$ and $\mathcal{N}_2(\Theta_2)$, where $\Theta_1$ and $\Theta_2$ are parameters. For the semi-supervised medical image segmentation task, the training dataset can be divided into two parts, labeled data $\mathcal{D}^l = \{(\mathcal{X}^l_k, \mathcal{Y}_k^l)\}_{i=1}^{A}$ and unlabeled data $\mathcal{D}^u = \{(\mathcal{X}_k^u)\}_{j=A+1}^{A+B}$ where $A \ll B$. Besides, the volume of 3D medical image and ground truth can be defined as $\mathcal{X} \in \mathbb{R}^{H\times W \times L}$, and $\mathcal{Y} \in \{0,1,...,K-1\}^{H\times W \times L}$ which indicates where targets and backgrounds are in $\mathcal{X}$.

\subsection{Backgoround of Evidential Deep Learning}
Dempster-Shafer evidence theory’s notion of belief mass over a frame of discernment as a Dirichlet Distribution is formalized by Subjective logic \cite{DBLP:books/sp/Josang16}. Therefore, it enables the application of evidence theory principles to quantify belief masses and uncertainty within a clearly defined theoretical framework. Specifically, in the field of classification task, the belief mass of each category, overall uncertainty mass and corresponding predicted probability can be given as \cite{DBLP:conf/nips/SensoyKK18}:
\begin{equation}
	\setlength\abovedisplayskip{5pt}
	b_n = e_n / S,\quad u = K / S, \quad p_n = \alpha_n / S,\quad \sum_{n=0}^{K-1}b_n+u=1
	\setlength\belowdisplayskip{-5pt}
\end{equation}
where $u \geq 0$, $b_n \geq 0$, $\alpha_n = e_n + 1$, $\boldsymbol{\alpha} = [\alpha_0,...,\alpha_{K-1})]$ and $S = \sum_{n=0}^{K-1}\alpha_n, n=0,..., K-1$. 

\subsection{Generalized Probabilistic Framework}
In this work, we extend EDL by integrating multiple objective sets in traditional evidence theory \cite{DBLP:series/sfsc/Dempster08a}. We map the inherent uncertainty of evidential deep learning onto the universal set corresponding to the discernment framework of traditional evidence theory \cite{DBLP:journals/pami/HanZFZ23}. This approach provides greater flexibility in the process of evidence-based feature fusion and uncertainty measurement. Specifically, let $b_{\mathcal{N}_1,l}^{(x,y,z)}$, $b_{\mathcal{N}_2,l}^{(x,y,z)}$, $b_{\mathcal{N}_1,u}^{(x,y,z)}$ and $b_{\mathcal{N}_2,u}^{(x,y,z)} \in \mathcal{R}^{C}$ represent the belief mass from the evidential classifier predictive results \cite{DBLP:conf/iclr/HanZFZ21} for a voxel at position $(x,y,z)$ by the segmentation output of sub-network $\mathcal{N}_1$ and $\mathcal{N}_2$ on labeled and unlabeled images respectively. Moreover, $K$ represents the number of classes. We define the basic probability mass assignments for a voxel as $\mathcal{M} = \{\{b^{(x,y,z)}_{(C_n)}\}_{n=0}^{K-1}, u^{(x,y,z)}\}$, with $u^{(x,y,z)} = 1 - \sum_{n=0}^{K-1}b^{(x,y,z)}_{(C_n)}$ signifying the original uncertainty. The generalized probability mass assignments which  maps the $u^{(x,y,z)}$ to multiple objective sets are then expressed as:
\begin{equation}
	\mathcal{M}^{(x,y,z)} = \{\{b^{(x,y,z)}_{(C_n)}\}_{n=0}^{K-1}, b^{(x,y,z)}_{(C_K)}\},\ C_K = \{C_0,...,C_{K-1}\}
\end{equation}
where $C_K$ denotes a multiple sets whose cardinality is $K-1$.

\subsection{Mutual Evidential Interaction for Pseudo-label Generation}
In the field of semi-supervised medical segmentation, most methods adopt a teacher-student network structure, where the teacher network is responsible for generating pseudo-labels for unlabeled data to guide the student network in the corresponding feature learning \cite{tarvainen2017mean}. However, some related work has abandoned this structure and instead uses two sub-networks to independently output pseudo-labels to guide the learning for unlabeled data, achieving quite excellent results \cite{wang2023mcf}. However, such a design may lead to potential problems like error accumulation and increased model bias. These issues could be caused by incorrect pseudo-labels generated by one of the networks, or when both networks have similar structures or biases, they may produce similar errors, leading to a lack of diversity in pseudo-labels \cite{zhang2021adaptive}. In these cases, the model may not be able to fully learn from the data, thus failing to achieve optimal performance. Therefore,  we propose a probability interaction strategy evolving from Dempster's combination rule \cite{DBLP:series/sfsc/Dempster08a}, where it allows the pseudo-label probability distributions generated by the two sub-networks to be reallocated and interact with each other, which can be defined as (the subscript $u$ is omitted for simplicity):
\begin{equation}\small
	\setlength\abovedisplayskip{5pt}
	\begin{aligned}
		&b_{(C_n)}^{(x,y,z)} = b^{(x,y,z)}_{\mathcal{N}_{1},(C_n)}b^{(x,y,z)}_{\mathcal{N}_{2},(C_n)} + \frac{|C_n|}{|C_n|+|C_N|}\\ &\cdot (b^{(x,y,z)}_{\mathcal{N}_{1},(C_n)}b^{(x,y,z)}_{\mathcal{N}_{2},(C_N)} + b^{(x,y,z)}_{\mathcal{N}_{2},(C_n)}b^{(x,y,z)}_{\mathcal{N}_{1}, (C_K)})
	\end{aligned}
	\label{eq2}
\end{equation}
where $b^{(x,y,z)}_{\mathcal{N}_{1},(C_n)}$ and $b^{(x,y,z)}_{\mathcal{N}_{2},(C_n)}$ represent prediction of sub-network $\mathcal{N}_1$ and $\mathcal{N}_2$ on unlabeled data for class $C_n$. Besides, the fused probability mass assignments are supposed to be normalized, $b_{(C_K)}^{(x,y,z)} = b^{(x,y,z)}_{\mathcal{N}_{1},(C_K)}b^{(x,y,z)}_{\mathcal{N}_{2},(C_K)}$, $|C_n| = 1$ and $|C_K| = K$. After the fusion of the predictions made by two sub-networks on unlabeled data, we can obtain the fused pseudo-labels for the unlabeled data. The significance of the coefficient $\frac{|C_k|}{|C_k|+|C_N|}$ is to prevent the values corresponding to the multiple objective set associated with uncertainty from being too trivial. We will also proceed to dynamically adjust the learning at the voxel level for the optimized uncertainty, which is illustrated in the next subsection. For contentious voxels, the corresponding pseudo-labels do not explicitly indicate the categories of voxel segmentation. Instead, the segmentation of these voxels is determined by the model learning the potential semantic connections among the surrounding voxels. Utilizing Eq. \ref{eq2}, we can integrate the probability distributions of pseudo-labels generated by the two sub-networks to obtain the final pseudo-labels for the unlabeled data. The predictions of the two networks for the unlabeled data undergo the same loss calculation process as in the training phase with labeled data. Unlike the training phase with labeled data, we consider the significant uncertainty that may exist in the pseudo-labeled data \cite{wang2021uncertainty}, and recognize that further reinforcing the model's learning of pseudo-label patterns may not aid in enhancing the model's segmentation performance on test data. Therefore, based on the integrated pseudo-labels, we propose using the uncertainty obtained during interaction for each voxel to compute its reliability. Prior to the loss calculation between pseudo-labels and predictions, we perform an operation on the pseudo-labels akin to an attention-based mask mechanism. Given the effectiveness of Shannon entropy \cite{DBLP:journals/bstj/Shannon48}, the definition of reliability synthesizing fused uncertainty for voxel at position $(x,y,z)$ can be given as:
\begin{equation}
	\mathcal{R}^{(x,y,z)} = e ^{b_{(C_K)}^{(x,y,z)} \sum_{n=0}^{N-1} (\boldsymbol{\zeta}_n log_2 \boldsymbol{\zeta}_n)}
\end{equation}
where $\boldsymbol{\zeta}_n$ represents $b_{(C_n)}^{(x,y,z)}$ and $e$ denotes natural constant. We adopt the same strategy as related work \cite{lu2023upcol}, utilizing reliability to mask the generated pseudo-labels, thereby further optimizing the quality of the pseudo-labels and reducing the likelihood of misleading the model during the learning process to generated the final pseudo-labels $\hat{y}$ for each voxel. 

\begin{figure*}[h]
	\centering
	\setlength{\belowcaptionskip}{-5pt}
	\includegraphics[scale=0.211]{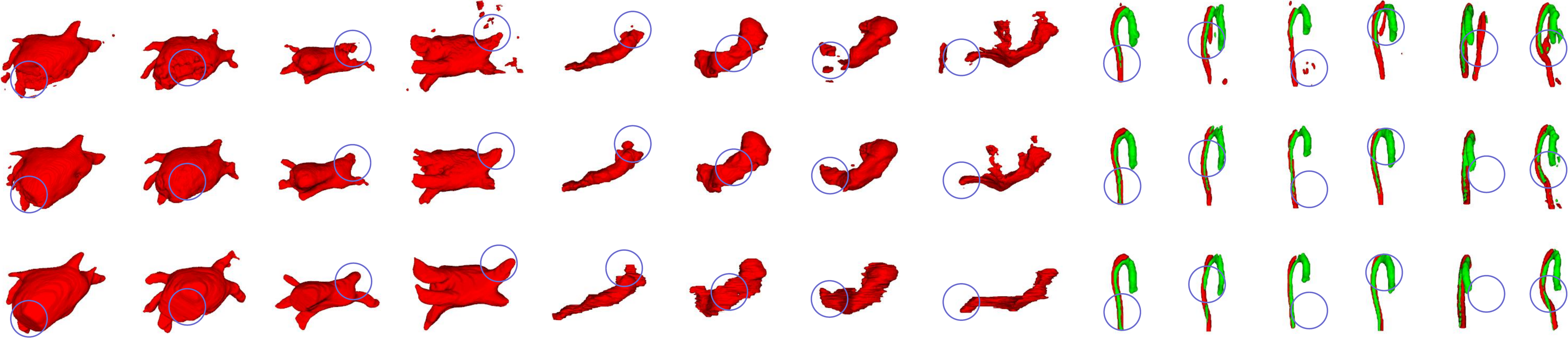}
	\label{figthree}
	\caption{Visualization results on the LA, Pancreas, and TBAD datasets. The first, second, and third rows show the results of the comparison method, the proposed method, and the ground truth (GT).}
\end{figure*}

\begin{table*}[htbp]\scriptsize
	\renewcommand{\arraystretch}{1.0251}
	\centering
	\caption{Comparisons with SOTA models on TBAD dataset in labeled ratio 5$\%$, 10$\%$ and 20$\%$}
	\setlength{\tabcolsep}{1.5mm}{
		\begin{tabular}{c |c c |c c c |c c c| c c c |c c c  }
			\hline
			\multicolumn{1}{c|}{\multirow{3}*{Model}}& \multicolumn{2}{c|}{\multirow{2}*{Scans Used}} &\multicolumn{12}{c}{Metrics}\\\cline{4-15}
			{}&{} &{} &\multicolumn{3}{c|}{Dice$\uparrow$}&\multicolumn{3}{c|}{Jaccard$\uparrow$}&\multicolumn{3}{c|}{95HD$\downarrow$}&\multicolumn{3}{c}{ASD$\downarrow$}  \\\cline{2-15}
			&{Labeled} & {Unlabled} & TL & FL & Mean & TL & FL & Mean & TL & FL & Mean& TL & FL & Mean \\ \cline{1-15}
			\multirow{2}*{V-Net}&(20\%)&(80\%)&55.51& 48.98 &52.25&39.81 &34.79& 37.30 &7.24& 10.17& 8.71& 1.27 &3.19 &2.23 \\
			&(100\%)&(0\%)&  75.98& 64.02& 70.00& 61.89& 50.05& 55.97& 3.16& 7.56& 5.36 &0.48& 2.44 &1.46 \\\hline
			MT \cite{DBLP:conf/iclr/TarvainenV17} &\multirow{5}*{(20\%)} & \multirow{5}*{(80\%)} &  57.62& 49.95& 53.78& 41.57& 35.52&38.54& 6.00& 8.98& 7.49& 0.97& 2.77& 1.87 \\
			UA-MT \cite{DBLP:conf/miccai/YuWLFH19}&&&70.91& 60.66 &65.78& 56.15& 46.24& 51.20& 4.44& 7.94& 6.19& 0.83& 2.37& 1.60  \\
			FUSSNet \cite{DBLP:conf/miccai/XiangQY22} &&&  79.73& 65.32& 72.53& 67.31& 51.74& 59.52& 3.46& 7.87& 5.67& 0.61& 2.93& 1.77  \\
			URPC \cite{DBLP:journals/mia/LuoWLCSCZMZ22} &&&  81.84& 69.15& 75.50& 70.35& 57.00& 63.68& 4.41& 9.13& 6.77& 0.93& \textcolor{red}{\textbf{1.11}} &1.02  \\
			UPCoL \cite{DBLP:conf/miccai/LuLQSZLY23} &&&  82.65& 69.74& 76.19& 71.49& 57.42& 64.45& 2.82& 6.81& 4.82& 0.43& 2.22& 1.33 \\ \hline
			\multirow{3}*{Ours}&(5\%)&(95\%)&81.83& \textcolor{red}{\textbf{73.51}} & \textcolor{red}{\textbf{77.67}} & 69.78&\textcolor{red}{\textbf{60.56}}&\textcolor{red}{\textbf{65.17}}&\textcolor{red}{\textbf{2.54}}&\textcolor{red}{\textbf{4.77}} & \textcolor{red}{\textbf{3.66}} & \textcolor{red}{\textbf{0.30}} & 1.32 & \textcolor{red}{\textbf{0.81}}\\&(10\%)&(90\%)&\textcolor{red}{\textbf{83.76}}& \textcolor{red}{\textbf{75.16}} & \textcolor{red}{\textbf{79.46}} & \textcolor{red}{\textbf{72.53}} &\textcolor{red}{\textbf{62.69}}&\textcolor{red}{\textbf{67.61}}&\textcolor{red}{\textbf{2.11}}&\textcolor{red}{\textbf{4.67}}&\textcolor{red}{\textbf{3.39}} & \textcolor{red}{\textbf{0.33}} & 1.32 & \textcolor{red}{\textbf{0.83}}\\
			&(20\%)&(80\%)&\textcolor{red}{\textbf{84.16}}& \textcolor{red}{\textbf{75.45}} & \textcolor{red}{\textbf{79.80}} & \textcolor{red}{\textbf{73.21}}&\textcolor{red}{\textbf{62.88}}&\textcolor{red}{\textbf{68.05}}&\textcolor{red}{\textbf{2.25}}&\textcolor{red}{\textbf{4.57}}&\textcolor{red}{\textbf{3.41}} & \textcolor{red}{\textbf{0.30}} & 1.22 & \textcolor{red}{\textbf{0.76}}\\
			\hline
	\end{tabular}}
	\label{4}
\end{table*}

\begin{table*}[h]\scriptsize
	\centering
	\renewcommand{\arraystretch}{1.055}
	\caption{Comparisons with SOTA models on three dataset in labeled ratio 5$\%$, 10$\%$ and 20$\%$}
	\setlength{\tabcolsep}{1.2mm}{\begin{tabular}{c | c | c c c c c c c c c c c c c c}
			\hline
			\multirow{14}*{\makecell{Pancreas-CT \\dataset}}&Labeled Ratio & Metrics  & UA-MT & SASSNet & DTC & URPC & MC-Net & SS-Net & Co-BioNet & BCP& A\&D & Ours\\
			\cline{2-13}
			&\multirow{4}*{(5\%)}   & Dice$\uparrow$ &  47.03&56.05&49.83&52.05& 54.99&56.35& 79.74& 80.33& 81.65&\textcolor{red}{\textbf{82.14}}\\
			&& Jaccard$\uparrow$ &  32.79& 41.56& 34.47&36.47&40.65&43.41& 65.66& 67.65& 69.11&\textcolor{red}{\textbf{69.86}}\\
			&& 95HD$\downarrow$ & 35.31&36.61& 41.16&34.02&16.03& 22.75&\textcolor{red}{\textbf{5.43}}& 11.78 &15.01 & 8.75\\
			&& ASD$\downarrow$ &  4.26&4.90& 16.53&13.16& 3.87& 5.39&\textcolor{red}{\textbf{2.79}}&4.32 &4.53 &2.91\\  \cline{2-13}
			&\multirow{4}*{(10\%)}   & Dice$\uparrow$ &66.96&66.69& 67.28& 64.73&69.07&67.40& 82.49& 81.54&82.25 &\textcolor{red}{\textbf{84.14}}\\
			&& Jaccard$\uparrow$ &  51.89& 51.66& 52.86&49.62& 54.36&53.06& 67.88&69.29&70.17 &\textcolor{red}{\textbf{72.86}}\\
			&& 95HD$\downarrow$ & 21.65 & 18.88& 17.74&21.90& 14.53&20.15& \textcolor{red}{\textbf{6.51}}&12.21&14.44 &9.90\\
			&& ASD$\downarrow$ &  6.25&5.76& \textcolor{red}{\textbf{1.97}}& 7.73& 2.28& 3.47& 3.26& 3.80&4.53 &2.72\\ \cline{2-13}
			&\multirow{4}*{(20\%)}   & Dice$\uparrow$ & 77.26 &77.66&78.27& 79.09& 78.17& 79.74&84.01&82.91 &82.56 &\textcolor{red}{\textbf{84.93}}\\
			&& Jaccard$\uparrow$ &  63.82&64.08& 64.75&65.99& 65.22& 65.42&70.00&70.97 &70.69 & \textcolor{red}{\textbf{73.95}} \\
			&& 95HD$\downarrow$ &  11.90&10.93& 8.36& 11.68& 6.90& 12.44&\textcolor{red}{\textbf{5.35}}&6.43 & 11.78 & 6.24\\
			&& ASD$\downarrow$ & 3.06& 3.05& 2.25& 3.31&\textcolor{red}{\textbf{1.55}}& 2.69&2.75& 2.25&3.42 &2.21\\ 
			\hline
			\multirow{14}*{\makecell{LA \\dataset}} & Labeled Ratio & Metrics  & UA-MT & SASSNet & DTC & URPC & MC-Net & SS-Net & Co-BioNet & BCP& A\&D & Ours\\
			\cline{2-13}
			&\multirow{4}*{(5\%)}   & Dice$\uparrow$   &82.26 & 81.60 & 81.25& 86.92&87.62&  86.33 &76.88 &88.02&89.93 & \textcolor{red}{\textbf{90.49}}\\
			&& Jaccard$\uparrow$ &70.98 & 69.63 & 69.33 & 77.03&78.25 &76.15 &66.76 &78.72&81.82  &\textcolor{red}{\textbf{82.75}} \\
			&& 95HD$\downarrow$ & 13.71&16.16 &14.90   &  11.13&  10.03&9.97& 19.09 &7.90&\textcolor{red}{\textbf{5.25}} & 5.95 \\
			&& ASD$\downarrow$ & 3.82&3.58&3.99& 2.28& 1.82& 2.31&  2.30 & 2.15 & 1.86
			&\textcolor{red}{\textbf{1.83}}\\ \cline{2-13}
			&\multirow{4}*{(10\%)}   & Dice$\uparrow$  & 87.79 & 87.54 &87.51&86.92& 87.62 & 88.55 &  89.20& 89.62 &90.31  &\textcolor{red}{\textbf{91.55}}\\
			&& Jaccard$\uparrow$ & 78.39 &  78.05 & 78.17& 77.03& 78.25 & 79.62& 80.68& 81.31 & 82.40 &\textcolor{red}{\textbf{84.46}}\\
			&& 95HD$\downarrow$ & 8.68 & 9.84&  8.23&11.13& 10.03 & 7.49&  6.44&  6.81 &\textcolor{red}{\textbf{5.58}} & 5.65\\
			&& ASD$\downarrow$ &  2.12 &  2.59& 2.36& 2.28 &  1.82 & 1.90&  1.90& 1.76 &1.64 & \textcolor{red}{\textbf{1.60}}\\ \cline{2-13}
			&\multirow{4}*{(20\%)}   & Dice$\uparrow$ & 88.88& 89.54& 89.42& 88.43& 90.12&  89.25&  91.26 &90.34 &90.42 &\textcolor{red}{\textbf{91.95}}\\
			&& Jaccard$\uparrow$ &80.21& 81.24&80.98 &81.15&82.12&81.62& 83.99&  82.50 &82.72 &\textcolor{red}{\textbf{85.14}}\\
			&& 95HD$\downarrow$ & 7.32& 8.24& 7.32& 8.21& 11.28&  6.45&  5.17& 6.75 &6.33 &\textcolor{red}{\textbf{5.11}}\\
			&& ASD$\downarrow$ &  2.26&1.99& 2.10&2.35&2.30& 1.80&  1.64&  1.77 &1.57 &\textcolor{red}{\textbf{1.37}}\\ \hline
			\multirow{14}*{\makecell{ACDC\\ dataset}}&Labeled Ratio & Metrics  & UA-MT & SASSNet & DTC & URPC & MC-Net & SS-Net & Co-BioNet & BCP& A\&D & Ours\\
			\cline{2-13}
			&\multirow{4}*{(5\%)}   & Dice$\uparrow$ &  46.04 & 57.77 & 56.90 & 55.87 & 62.85 & 65.83 & 87.46 & 87.59& 86.51 & \textcolor{red}{\textbf{89.60}}\\
			&& Jaccard$\uparrow$ & 35.97 & 46.14 & 45.67 & 44.64 & 52.29 & 55.38 &77.93 & 78.67& 76.61 &  \textcolor{red}{\textbf{81.67}} \\
			&& 95HD$\downarrow$ & 20.08 & 20.05 & 23.36 & 13.60 & 7.62 & 6.67 &\textcolor{red}{\textbf{1.11}} &1.90 &2.13 & 4.80\\
			&& ASD$\downarrow$ & 7.75 & 6.06 & 7.39 & 3.74 & 2.33 & 2.28 & \textcolor{red}{\textbf{0.41}} & 0.67 & 0.84 & 1.31 \\  \cline{2-13}
			&\multirow{4}*{(10\%)}  & Dice$\uparrow$ & 81.65 & 84.50 & 84.29 & 83.10 & 86.44 & 86.78 & 88.49& 88.84&88.12& \textcolor{red}{\textbf{90.40}}\\
			&& Jaccard$\uparrow$ & 70.64 & 74.34 & 73.92 & 72.41 & 77.04 & 77.67 & 79.76 & 80.62& 79.39 & \textcolor{red}{\textbf{82.94}} \\
			&& 95HD$\downarrow$ & 6.88 & 5.42 & 12.81 & 4.84 & 5.50 & 6.07 & 3.70 & 3.98& 13.03 & \textcolor{red}{\textbf{3.38}} \\
			&& ASD$\downarrow$ &2.02 & 1.86 & 4.01 & 1.53 & 1.84 & 1.40 &1.14 & 1.17& 3.21 & \textcolor{red}{\textbf{0.83}}\\ \cline{2-13}
			&\multirow{4}*{(20\%)} & Dice$\uparrow$ & 85.61 & 86.45 & 87.10 & 85.44 & 87.04 & 87.41 & 89.51& 89.12&88.85 & \textcolor{red}{\textbf{90.95}} \\
			&& Jaccard$\uparrow$ & 75.49 & 77.20 & 78.15 & 76.36 & 78.01 & 78.82 &81.64 &81.03 & 80.62 & \textcolor{red}{\textbf{83.80}} \\
			&& 95HD$\downarrow$ & 5.91 & 6.63 & 6.76 & 5.93 & 5.35 & 4.79 & 4.72& \textcolor{red}{\textbf{3.40}}& 4.26 & 3.84\\
			&& ASD$\downarrow$ & 1.79 & 1.98 & 1.99 & 1.70 & 1.67 & 1.48& 1.52&0.97 &1.39 &\textcolor{red}{\textbf{0.85}} \\ \hline
	\end{tabular}}
	\label{2}
\end{table*}

\subsection{Asymptotic Fisher Evidential Deep Learning}
We propose optimizing the model labeled learning process by utilizing the way of uncertainties estimation designed for unlabeled learning part and the model's voxel-wise evidential predictions of labeled data. For learning with labeled data, we find that integrating the predictions of the two sub-networks in the same manner as with pseudo-labels, results in certain performance gains. The measurement of uncertainty for each voxel in the labeled data learning phase is the weighted mean of the original and fused uncertainty values, which can be given as (the subscript $l$ is omitted for simplicity):
\begin{equation}
	b_{\mathcal{N}_i,C_K}^{final} = \lambda_1 b_{\mathcal{N}_i, C_K}^{(x,y,z)} + \lambda_2 b_{(C_K)}^{(x,y,z)}, \quad \lambda_1
	+ \lambda_2 = 1
	\end{equation}
where $i = 1, 2$, $b_{\mathcal{N}_i,C_K}^{final}$ and $b_{(C_K)}^{(x,y,z)}$ represent final uncertainty measure and the one from combined evidence for labeled data voxel at position $(x,y,z)$. Based on the optimized uncertainty measurement of voxel from sub-networks on labeled data and original uncertainty measure $b_{(C_K)}^{(x,y,z)}$ in pseudo-labels for unlabeled data from evidential predictive results, we propose to evaluate uncertainty of the model's prediction and pseudo-labels for each voxel in labeled and unlabeled data to guide the model for targeted learning. We aim for the model to initially focus on voxels and pseudo-labels with smaller uncertainties, and then gradually shift towards understanding features of voxels with higher uncertainties and the ones corresponding to more unreliable pseudo-labels  as the training progresses. Drawing inspiration from curriculum learning \cite{bengio2009curriculum}, the proposed approach enhances the learning of complex information, building on a firm grasp of simpler characteristics. The voxel-level weight adjustment function is defined as:
\begin{equation}\small
	\setlength\abovedisplayskip{8pt}
	\setlength\belowdisplayskip{8pt}
	\omega(\boldsymbol{q,v}) = \varXi \cdot Tanh(\psi(h(\boldsymbol{v}))\zeta(\boldsymbol{q})) + 1
\end{equation}
where $\varXi$ is devised to control the change amplitude of dynamic weights. $\psi(h(\boldsymbol{v})) = \frac{2h(\boldsymbol{v})}{V}-1\in [-1,1], v = 1,...,V$, $h(v)$ represents the rank number of voxel $v$ through sorting uncertainties of voxels in labeled data and pseudo-labels in a descending order. Besides, $\zeta(\boldsymbol{q}) = \frac{2q}{Q}-1 \in [-1,1], q=1,...,Q$, where $Q$ denotes the total number of total training epochs and $q$ is the current epoch index. For unlabeled data, considering the importance of modeling uncertainties contained in classes, we propose to integrate the voxel-level weight for pseudo-labels with the KL term-free optimization objective presented by fisher information-based evidential deep learning $\mathcal{I}$-EDL \cite{DBLP:conf/icml/DengCYLH23} for a more robust learning pattern on voxel features from unlabeled data \cite{DBLP:conf/aaai/YuDL0JCH24}, which can be defined as (the subscript $C_n$ is replaced by $n$ and superscript ${(x,y,z)}$ is omitted):
\begin{equation}\scriptsize
	\begin{split}&
\mathcal{L}_{j,\boldsymbol{v}}^{\mathcal{I}} = \omega(\boldsymbol{q,v})(\sum_{n=0}^{K-1} ( (\hat{y}_{\boldsymbol{v}n} - \frac{\alpha_{\boldsymbol{v}n}}{S_{\boldsymbol{v}}})^2 + \frac{\alpha_{\boldsymbol{v}n}(S_{\boldsymbol{v}} - \alpha_{\boldsymbol{v}n})}{S_{\boldsymbol{v}}^2(S_{\boldsymbol{v}} + 1)})\\&\qquad \qquad \qquad\quad\psi_1(\alpha_{\boldsymbol{v}n}) - \lambda_2 \log |\mathcal{I}(\boldsymbol{\alpha}_{\boldsymbol{v}})|),
	\end{split}
\end{equation}
where $\hat{y}_{\boldsymbol{v}}$ represents one-hot encoded ground-truth of $v_{th}$ voxel and $\mathcal{L}_{j,\boldsymbol{v}}^{fiel}$ denotes the loss between $v_{th}$ voxel in $j_{th}$ unlabeled sample and corresponding integrated pseudo-labels. 

\begin{figure}[htbp]
	\centering
	\setlength{\belowcaptionskip}{-15pt}
	\includegraphics[scale=0.2]{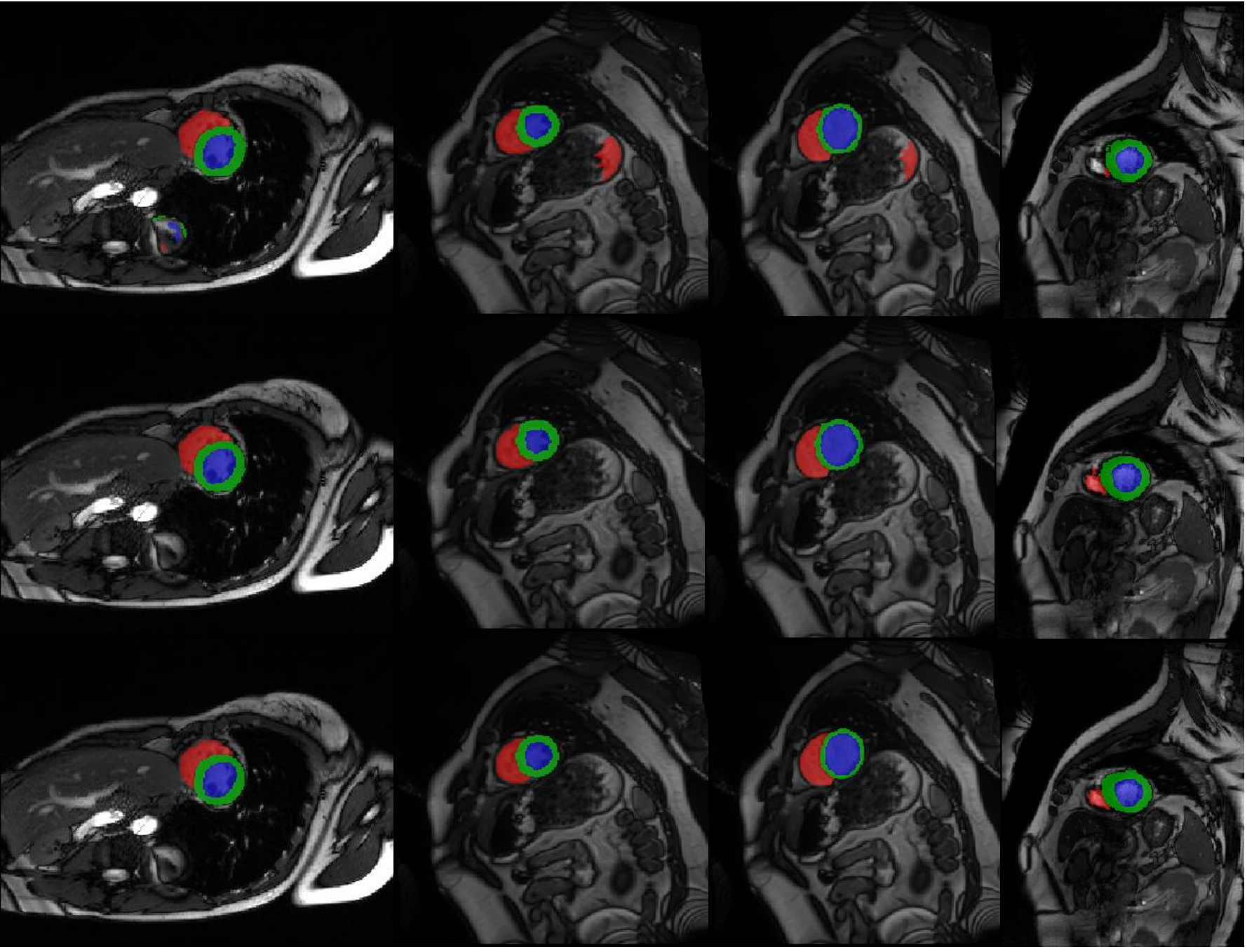}
	\caption{Visualization results on the ACDC dataset. The first, second, and third rows are the results of the comparison method, the proposed method, and the ground truth (GT).}
	\label{figthreeeeeee}
\end{figure}

\subsection{Optimization Objective}
Update of model parameters is divided into learning from labeled and unlabeled data \cite{wang2023mcf,DBLP:conf/miccai/LuLQSZLY23}. For labeled data, the sub-networks' predictions are compared with the labels to compute Dice and Cross-Entropy loss, resulting in $\mathcal{L}_l^{\mathcal{N}_1}$ and $\mathcal{L}_l^{\mathcal{N}_2}$. Utilizing the both kinds of uncertainties from predictions of labeled data and generated pseudo-labels, we can apply weighting to the loss calculation for each voxel of labeled data and unlabeled data. The process generating asymptotic loss for labeled and unlabeled data can be given as:
\begin{equation}
	\mathcal{L}_{w,l}^{\epsilon} = \sum_{i=1}^{A}\sum_{\boldsymbol{v}_l=1}^{V}\omega(\boldsymbol{q},\boldsymbol{v}_l)\mathcal{L}^{\epsilon}_{i,\boldsymbol{v}_l} / V, \quad \mathcal{L}_u^{\epsilon,\mathcal{I}} = \sum_{j=1}^{B}\sum_{\boldsymbol{v}_u=1}^{V'} \mathcal{L}_{j,\boldsymbol{v}_u}^{\mathcal{I}} / V'
\end{equation}
where $\epsilon = \mathcal{N}_1, \mathcal{N}_2$, $V$ and $V'$ represent a total number of voxels in a labeled and unlabeled sample. Besides, $\mathcal{L}^{\epsilon}_{i,\boldsymbol{v}}$ denotes the loss of $\boldsymbol{v}_{th}$ voxel in $i_{th}$ sample and $h(\boldsymbol{v})$ denotes corresponding ordinal number of voxel in $i_{th}$ sample. For the unlabeled data part, the prediction of each sub-network is compared with the optimized generated pseudo-labels to compute the losses consisting of Dice and Cross-Entropy loss, denoted as $\mathcal{L}_{u}^{\mathcal{N}_1}$ and $\mathcal{L}_{u}^{\mathcal{N}_2}$. For the update of each sub-network's parameters, the final optimization objective can be given as:
\begin{equation}
	\mathcal{L}^{\epsilon}_{medl} = \mathcal{L}_{l}^{\epsilon} + \mathcal{L}_{u}^{\epsilon}  + \mathcal{L}_{w,l}^{\epsilon}+ \lambda_{GWU}\mathcal{L}_u^{\epsilon,\mathcal{I}}
\end{equation}
where $\lambda_{GWU}$ are self-adaptive hyper-parameters.

\section{Experiments}
\label{sec:ex}

In our framework, subnet $\mathcal{N}_1$ is based on the VNet architecture, which is widely recognized for its effectiveness in medical image segmentation. To facilitate error correction between subnets, it's essential that the performance disparity among the diverse subnets is minimal. To this end, we have substituted VNet's encoder with a 3D ResNet34, creating a second subnet $\mathcal{N}_2$ known as 3D-ResVNet. When processing the data for final output, the system computes the mean of both subnet outputs. PyTorch is employed to run this system, leveraging the computational power of an NVIDIA 4090 GPU. Our parameter configurations align well with established benchmarks for comparison. We employ an SGD optimizer to modify network parameters, setting the weight decay to 0.0001 and the momentum to 0.9. The learning rate starts at 0.01, reducing by a factor of 10 at intervals of 2500 iterations, up to 6000 iterations in total. We utilize a batch size of four, comprising two labeled and two unlabeled data volumes. Following \cite{DBLP:conf/miccai/LiZH20}, Gaussian warming up function is utilized to control the weight {\small $\lambda_{GWU}$}.

\begin{table}[htbp]\tiny
	\renewcommand{\arraystretch}{1.1}
	\centering
	\caption{Comparison with other methods on the ACDC test set. DSC (\%) and ASSD (mm) are reported with selected 14 labeled scans and 126 unlabeled scans \cite{DBLP:journals/corr/abs-2404-07032} for semi-supervised training. The bold font represents the best performance.}
	\setlength{\tabcolsep}{1.4mm}{
		\begin{tabular}{l|c|cccc|cccc}
			\hline
			\multirow{2}*{Method} & \textbf{Scans} & \multicolumn{4}{c|}{RV} & \multicolumn{4}{c}{Myo} \\
			\cline{3-10}
			& L/U & DSC $\uparrow$ & JAC $\uparrow$ & ASD $\downarrow$ & 95HD $\downarrow$ & DSC $\uparrow$ & JAC $\uparrow$ & ASD $\downarrow$ & 95HD $\downarrow$ \\
			\hline
			CPS \cite{DBLP:conf/cvpr/ChenYZ021} & 14/126 & 83.51 & 71.30 & \textcolor{red}{\textbf{0.46}} & 4.20 & 85.91 & 74.64 & 1.17 & 5.34 \\
			ASE-Net \cite{DBLP:journals/tmi/LeiZDWWN23} & 14/126 & 83.05 & 72.72 & \textcolor{red}{\textbf{0.46}} & 4.20 & 85.61 & 75.18 & 1.17 & 5.34 \\
			CoraNet \cite{shi2022inconsistency} & 14/126 & 83.05 & 72.72 & \textcolor{red}{\textbf{0.46}} & 4.20 & 85.61 & 75.18 & 1.17 & 5.34 \\
			MCF \cite{wang2023mcf} & 14/126 & 83.76 & 74.16 & 1.53 & 2.74 & 85.40 & 75.18 & 1.08 & 6.35 \\
			ETC-Net \cite{DBLP:journals/corr/abs-2404-07032} & 14/126 & 86.52 & 77.28 & 1.48 & 5.52 & 85.66 & 75.26 & \textcolor{red}{\textbf{0.82}} & 3.50 \\
			Ours & 14/126 & \textcolor{red}{\textbf{89.01}} & \textcolor{red}{\textbf{80.80}} & 7.28 & \textcolor{red}{\textbf{1.70}} & \textcolor{red}{\textbf{86.43}} & \textcolor{red}{\textbf{76.40}} & 2.87 & \textcolor{red}{\textbf{0.98}} \\
			\hline
			\multirow{2}*{Method} & \textbf{Scans} & \multicolumn{4}{c|}{LV} & \multicolumn{4}{c}{Avg} \\
			\cline{3-10}
			& L/U & DSC $\uparrow$ & JAC $\uparrow$ & ASD $\downarrow$ & 95HD $\downarrow$ & DSC $\uparrow$ & JAC $\uparrow$ & ASD $\downarrow$ & 95HD $\downarrow$ \\
			\hline
			CPS \cite{DBLP:conf/cvpr/ChenYZ021} & 14/126 & 89.86 & 80.28 & 2.14 & 6.42 & 86.43 & 75.40 & 1.26 & 5.32 \\
			ASE-Net \cite{DBLP:journals/tmi/LeiZDWWN23} & 14/126 & 89.86 & 80.28 & 2.14 & 6.42 & 86.18 & 76.06 & 1.26 & 5.32 \\
			CoraNet \cite{shi2022inconsistency} & 14/126 & 89.86 & 80.28 & 2.14 & 6.42 & 86.18 & 76.06 & 1.26 & 5.32 \\
			MCF \cite{wang2023mcf} & 14/126 & 89.15 & 81.48 & 2.35 & 7.08 & 86.77 & 76.27 & 1.62 & 5.39 \\
			ETC-Net \cite{DBLP:journals/corr/abs-2404-07032} & 14/126 & 92.07 & 85.68 &\textcolor{red}{\textbf{1.43}} & 4.25 & 88.08 & 79.40 & \textcolor{red}{\textbf{1.24}} & 4.42 \\
			Ours & 14/126 & \textcolor{red}{\textbf{93.34}} & \textcolor{red}{\textbf{87.81}} & 4.25 & \textcolor{red}{\textbf{1.23}} & \textcolor{red}{\textbf{90.40}} & \textcolor{red}{\textbf{82.94}} & 3.38 &\textcolor{red}{\textbf{0.83}} \\
			\hline
		\end{tabular}
	}
	\label{acdc}
\end{table}

\subsection{Datasets Description}
$\!\!\!\!\!\!$\textbf{The Left Atrium Dataset (LA). } The LA dataset \cite{DBLP:journals/mia/XiongXHHBZVRMYH21} consists of 100 3D gadolinium-enhanced MR imaging volumes. Its spatial resolution is 0.625 × 0.625 × 0.625{\scriptsize \(mm^3\)}. The dataset is divided into 5 folds, each with 20 volumes. For the stage of preprocessing, all volumes are normalized to zero mean and unit variance, and then the extensive edges of each 3D MRI volume are cropped according to the targets. In the process of training, we randomly crop the training volume to 112 × 112 × 80 and use it as the input. For the inference stage, we utilize the same size sliding window and a stride of 18 × 18 × 4 to obtain the final segmentation results. 
\\ \textbf{The Pancreas-CT Dataset. } Pancreas-CT dataset \cite{DBLP:conf/miccai/RothLFSLTS15} contains 82 contrastk-enhanced abdominal annotated 3D CT volumes. The size of each CT volume is 512 × 512 × D where D ranges from 181 to 466. The Pancreas-CT dataset is divided into four folds and the number corresponding to each fold is 20, 20, 21, 21. For the stage of preprocessing, a soft tissue \cite{DBLP:conf/aaai/LuoCSW21} CT window of [-120,240] HU is utilized and the CT scan is cropped centered on the pancreatic region while the extensive edges of 25 voxels are magnified. We crop the training volume to 96 × 96 × 96 randomly, with a stride of 16 × 16 × 16 at the inference stage.\\ 
\textbf{TBAD Dataset.} The Multi-center type B aortic dissection (TBAD) dataset comprises 124 CTA scans \cite{yao2021imagetbad}. Experienced radiologists meticulously label the TBAD dataset, dividing it into 100 training scans and 24 test scans. This dataset includes both publicly accessible data and contributions from UPCoL. It standardizes to a resolution of 1mm$^3$ and resizes to dimensions of 128×128×128. In each dataset, a maximum of 20\% of the labeled training data is used, with voxel intensities normalized to achieve zero mean and unit variance.\\
\textbf{ACDC Dataset.} The Automated Cardiac Diagnosis Challenge dataset \cite{bernard2018deep} encompasses 200 scans of 100 patients, which belong to four classes: background, right ventricle, left ventricle, and myocardium. We select 100 scans  \cite{DBLP:conf/cvpr/BaiCL0023} and split them into 70\% for training, 10\% for validation, and 20\% for testing for experiments in Table \ref{2}. A 2D U-Net serves as the backbone, accepting inputs of size $256\times 256$. The zero-value region of the mask $\mathcal{M}$ has a dimension of $170\times 170$.\\
\textbf{BraTs2018 dataset.} The BraTs2018 brain tumor dataset comprises 285 patient cases for training and 66 for validation. Each patient has four types of brain MRI volumes: Flair, T1, T2, and T1ce, along with the tumor segmentation ground truth. Each volume is sized at 155 × 240 × 240 voxels. To facilitate the training process, we first sliced the 3D volumes into 2D images and normalized each modality for the brain region using the mean and standard deviation. Subsequently, the images were center-cropped to 160 × 160 pixels.\\
\textbf{Metrics. }There are four metrics which are used for evaluating our proposed method: edge-sensitive indicators: 95\% Hausdorff Distance (95HD) and Average Surface Distance (ASD);  regional-sensitive metrics: Dice similarity coefficient (Dice) and Jaccard similarity coefficient (Jaccard).\\

\subsection{Comparison on the four benchmark dataset}
We compare the performance of the proposed model with previous SOTA methods including UA-MT \cite{DBLP:conf/miccai/YuWLFH19}, SASSNet \cite{DBLP:conf/miccai/LiZH20}, DTC \cite{DBLP:conf/aaai/LuoCSW21}, MC-Net \cite{DBLP:journals/mia/WuGZXZXC22}, SS-Net \cite{DBLP:conf/miccai/WuWWGC22}, Co-BioNet \cite{DBLP:journals/natmi/PeirisHCEH23}, BCP \cite{DBLP:conf/cvpr/BaiCL0023} and A\& D \cite{DBLP:conf/nips/WangL23} on Pancreas-CT, LA and ACDC datasets. Besides, the performance of our model is also compared with methods like MT, UA-MT, FUSSNet \cite{DBLP:conf/miccai/XiangQY22}, URPC \cite{DBLP:journals/mia/LuoWLCSCZMZ22} and UPCoL \cite{DBLP:conf/miccai/LuLQSZLY23} on TBAD dataset. This paper presents experiments conducted on five datasets and the results are provided in Table \ref{4} and \ref{2}. For the TBAD dataset, the proposed method achieves superior performance using only 5\% of labeled data compared to other methods using 20\% of labeled data. With 10\% and 20\% of labeled data, the proposed approach demonstrates significant performance improvements. For the remaining datasets, the proposed method still outperforms others in most cases demonstrating its effectiveness.

\begin{table}[ht]\scriptsize
	\centering
	\renewcommand{\arraystretch}{1.03}
	\caption{The comparison of different methods on ACDC dataset on different semi-supervised labeled data ratio settings.}
	\setlength{\tabcolsep}{1.4mm}{\begin{tabular}{l|ccc|ccc}
			\hline
			\multirow{2}*{Method} & \multicolumn{3}{c|}{10\%} & \multicolumn{3}{c}{20\%} \\\cline{2-7}
			&DSC \scalebox{0.75}{$\uparrow$} & 95HD \scalebox{0.75}{$\downarrow$} & ASD \scalebox{0.75}{$\downarrow$} & DSC \scalebox{0.75}{$\uparrow$} & 95HD \scalebox{0.75}{$\downarrow$} & ASD \scalebox{0.75}{$\downarrow$}\\
			\hline
			ICT \cite{DBLP:conf/ijcai/VermaLKBL19} & 83.54 & 8.42 & 2.46 & 86.52 & 5.65 & 1.90  \\
			CPS \cite{DBLP:conf/cvpr/ChenYZ021} & 84.70 & 8.25 & 2.35 & 87.47 & 5.98 & 1.24  \\
			URPC \cite{DBLP:journals/mia/LuoWLCSCZMZ22} & 82.07 & 5.62 & 1.88 & 85.87 & 5.71 & 1.75 \\
			EVIL \cite{DBLP:conf/isbi/ChenYSWQZ23} & 85.91 & 3.91 & 1.36 & 88.22 & 4.01 & 1.21 \\
			Ours & \textcolor{red}{\textbf{90.40}} & \textcolor{red}{\textbf{3.38}} & \textcolor{red}{\textbf{0.83}} & \textcolor{red}{\textbf{90.95}} & \textcolor{red}{\textbf{3.84}} & \textcolor{red}{\textbf{0.85}}\\
			\hline
	\end{tabular}}
	\label{11111}
\end{table}

\begin{table}[ht]\footnotesize
	\centering
	\renewcommand{\arraystretch}{1.}
	\caption{Comparison with the state-of-the art methods (Dice score) on the BraTS 2018 validation dataset.}
	\setlength{\tabcolsep}{2.2mm}{\begin{tabular}{lcccc}
			\hline
			\multirow{2}*{Methods} & \multicolumn{3}{c}{Dice} & Dice \\\cline{2-5}
			& WT & TC & ET & mean \\
			\hline
			ELUnet \cite{DBLP:conf/isbi/HuangRD21} & 86.16 &  90.27 & 85.15 & 87.19 \\
			No-new-Net \cite{DBLP:conf/miccai/IsenseeKWBM18} & 90.62 & 84.54 & 80.12 & 85.09 \\
			DMFNet \cite{DBLP:conf/miccai/0001LDZL19} & \textcolor{red}{\textbf{91.26}} & 86.34 & 80.87 & 86.15 \\
			NVDLMED \cite{DBLP:conf/miccai/Myronenko18} & 90.68 & 86.02 & 81.73 & 86.14 \\
			C-VNet \cite{DBLP:journals/prl/SharifLKS20} & 90.48 & 83.64 & 77.68 & 83.93 \\
			Ours & 90.78 & \textcolor{red}{\textbf{90.59}} & \textcolor{red}{\textbf{86.24}} & \textcolor{red}{\textbf{89.20}} \\ 
			\hline
	\end{tabular}}
	\label{22222}
\end{table}

\subsection{Ablation Studies}

In this section, we will explore the contributions of various components to the overall performance of the proposed method and regard MCF \cite{wang2023mcf} as baseline. CAEF (U) denotes utilizing evidential fusion to generate the pseudo-labels for unlabeled data. CAEF (L) denotes the fusion of evidential predictions from different networks for labeled data. FIE denotes fisher evidential deep learning. Besides, UMAL (U) and UMAL (L) denote the uncertainty measure-based asymptotic learning for unlabeled data and labeled data, respectively. To be specific, each proposed component will sequentially be added to the baseline. Table \ref{table:component_ablation}. provides the experimental results. According to the table, it can be observed that both EF (U) and EF (L) bring remarkable performance improvement (an increase of 3.05\%, 4.27\% and 6.1 on metrics of Dice, Jaccard and ASD). Subsequently, when the components of UMAL (U) and UMAL (L) are integrated into the baseline method, the performance also appears great improvements (The Dice score increases to 91.71 and the Jaccard score to 84.75, with 95HD and ASD reducing to 5.30 and 1.42). It can be proved that the uncertainty measure-based asymptotic learning is capable of boosting performance. Finally, when the fisher evidential deep learning is integrated into baseline, the model yields the best performance. Additionally, the impact of different $\lambda$ values on model's performance is also explored. Table \ref{table:lambda_ablation}. demonstrates the experimental results. According to the table, based on the Dice and Jaccard indices, $\lambda_1, \lambda_2 = (0.5, 0.5)$ appears to be the optimal value, providing the best overlap measures. However, there is a clear trade-off between overlap measures (Dice and Jaccard) and boundary accuracy measures (95HD and ASD).

\begin{table} \footnotesize
	\centering
	\setlength{\belowcaptionskip}{-5pt}
	\caption{Component Ablation Study of the MEDL.}
	\renewcommand{\arraystretch}{0.953}
	\setlength{\tabcolsep}{1.4mm}{
		\begin{tabular}{lcccc}
			\hline
			\textbf{Method}  & Dice$\uparrow$ & Jaccard$\uparrow$ & 95HD$\downarrow$ & ASD$\downarrow$   \\
			\hline
			Baseline &  87.06 & 77.83& \textcolor{red}{\textbf{2.67}}& 7.81 \\
			+ CAEF (U)  &  89.04 & 80.37 & 7.62 & 1.95 \\
			+ CAEF (L)  & 90.11 & 82.10 & 6.47 & 1.71  \\
			+  UMAL (U)  &90.71 & 83.14 & 6.02 & 1.52  \\
			+  UMAL (L) & 91.71 & 84.75 & 5.30 & 1.42 \\
			+  FIE &  \textcolor{red}{\textbf{91.95}} & \textcolor{red}{\textbf{85.14}} & 5.11 & \textcolor{red}{\textbf{1.37}} \\
			\hline
	\end{tabular}}
	\label{table:component_ablation}
\end{table}

\begin{table} 
	\centering
	\caption{Ablation Study of Varying $\lambda$ Values.}
	\renewcommand{\arraystretch}{0.96}
	\setlength{\tabcolsep}{2.2mm}{
		\begin{tabular}{ccccc}
			\hline
			$\lambda_1$,  $\lambda_2$& Dice$\uparrow$ & Jaccard$\uparrow$ & 95HD$\downarrow$ & ASD$\downarrow$ \\
			\hline
			(0.3, 0.7) &90.37 & 82.88 &  \textcolor{red}{\textbf{3.28}} & 0.86   \\
			(0.4, 0.6) &90.31 & 82.85 & 3.75 &  \textcolor{red}{\textbf{0.79}}\\
			(0.5, 0.5)&\textcolor{red}{\textbf{90.95}} & \textcolor{red}{\textbf{83.80}} &3.84 &0.85   \\
			(0.6, 0.4) & 90.77 & 83.56 & 3.68 & 0.81   \\
			(0.7, 0.3) & 90.70 & 83.45 & 3.82 & 0.87\\
			\hline
	\end{tabular}}
	\label{table:lambda_ablation}
\end{table}

\subsection{Differences between the MEDL and EDL-based solutions}
\begin{figure} 
	\centering
	\setlength{\belowcaptionskip}{-0.4cm}
	\includegraphics[scale=0.23]{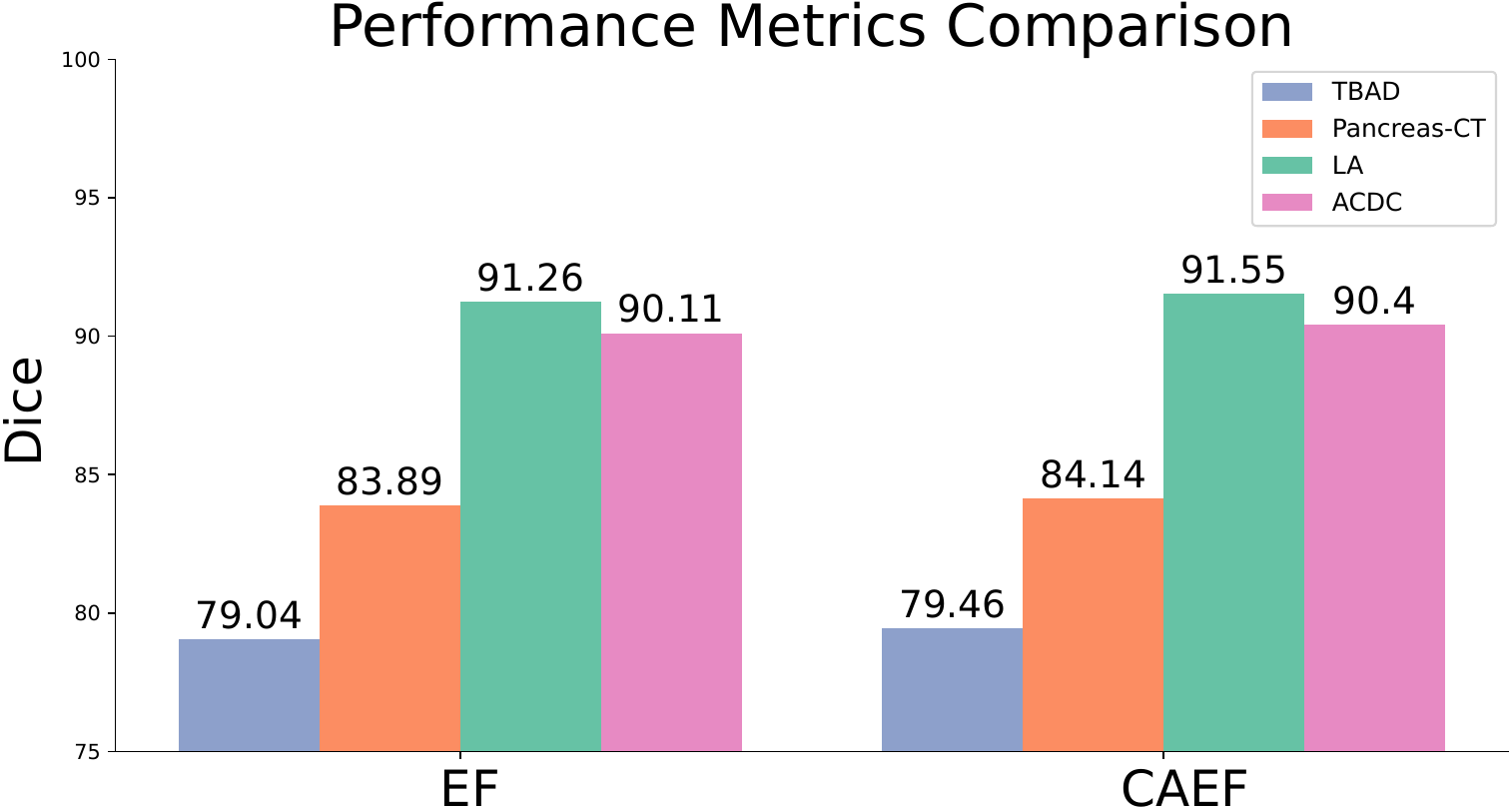}
	\caption{10\% labeled ratio performance comparison on the usage of EF and CAEF.}
	\label{PCEC}
\end{figure}
Several studies have applied EDL to semi-supervised medical image segmentation. Nevertheless, our work fundamentally differs from ETC-Net \cite{DBLP:journals/corr/abs-2404-07032}, EVIL \cite{DBLP:conf/isbi/ChenYSWQZ23} (EL), and ELUnet \cite{DBLP:conf/isbi/HuangRD21} in key aspects (perforamnce comparison is provided in Table \ref{acdc}, \ref{11111} and \ref{22222}): EL adopts the original EDL loss and probability definitions, substituting these into the dice loss, whereas we propose using belief entropy to re-evaluate voxel-level evidence, tightly integrating EDL loss with uncertainty to reduce feature confusion. While ETC-Net and ELUnet use evidential fusion (EF) for generating pseudo labels and combining predictions, we utilize an improved class-aware evidential fusion (CAEF) formula to integrate pseudo labels from different network architectures as also evidenced by recent work \cite{wang2024evidential}, and combine predictions from both networks on labeled data enhancing robustness and predictive performance. 
\subsection{Ablation study of EF and CAEF}
Furthermore, we use a class-aware evidential fusion strategy that adjusts according to the number of segmentation classes corresponding to each voxel. The performance comparison between conventional evidential fusion (EF) and CAEF is shown in Figure \ref{PCEC}. It can be concluded that CAEF provides better performance compared to EF. We argue this is because, as the total number of classes corresponding to the voxels increases, the interaction between confidence and uncertainty associated with the voxels should be suppressed during the fusion process. Without discounting this process, uncertain information is likely to dominate, potentially leading to confusion in some scenarios and performance degradation.

\section{Conclusion}
\label{sec:con}
The proposed MEDL framework effectively enables models to assess the uncertainty of both labeled and unlabeled data. Through class-aware evidential fusion, we achieve the synthesis of pseudo-labels generated by heterogeneous network models while integrating their evidential predictions for labeled data, resulting in more robust predictive results. The uncertainty-based asymptotic learning strategy further enhances the model's ability to learn reliable and manageable features, asymptotically shifting attention to more challenging features as training progresses. Moreover, extensive experimental results on five mainstream medical segmentation datasets strongly validate the effectiveness of MEDL.

\bibliographystyle{IEEEtran}
\bibliography{main2.bib}
\end{document}